\documentclass[aps,twocolumn,prl]{revtex4}
     
\usepackage{graphicx}
\usepackage{dcolumn}
\usepackage{bm}

\begin{document}
\preprint{APS/123-QED}

\title{Quasiparticle entropy in the high-field superconducting phase of CeCoIn$_5$}

\author{Y. Tokiwa$^1$}
\author{E.D. Bauer$^2$}
\author{P. Gegenwart$^1$}

\affiliation{$^1$I. Physikalisches Institut,
Georg-August-Universit\"{a}t G\"{o}ttingen, 37077 G\"{o}ttingen,
Germany}
\affiliation{$^2$Los Alamos National Laboratory, Los Alamos, New Mexico 87545, USA}

\date{\today}

\begin{abstract}
The heavy-fermion superconductor CeCoIn$_5$ displays an additional transition within its superconducting (SC) state, whose nature is characterized by high-precision studies of the isothermal field dependence of the entropy, derived from combined specific heat and magnetocaloric effect measurements at temperatures $T\geq 100$~mK and fields $H\leq 12$~T aligned along different directions. For any of these conditions, we do not observe an additional entropy contribution upon tuning at constant temperature by magnetic field from the homogeneous SC into the presumed Fulde-Ferrell-Larkin-Ovchinnikov (FFLO) SC state. By contrast, for $H\parallel [100]$ a reduction of entropy was found which quantitatively agrees with the expectation for 
spin-density-wave (SDW) order without FFLO superconductivity. Our data exclude the formation of a FFLO state in CeCoIn5 for out-of-plane field directions, where no SDW order exists. 
\end{abstract}

\maketitle
Superconductivity can be affected by an imbalance of density of states (DOS) for spin-up and spin-down electrons introduced by Zeeman splitting under magnetic field, leading to a finite total momentum Cooper pairing, 
in contrast to the usual BCS pairing.
Since Fulde, Ferrell, Larkin and Ovchinnikov~\cite{fulde-ferrell:pr-64,larkin-ovchinnikov:sov-phys65} (FFLO) have predicted such a superconducting (SC) state more than forty years ago, numerous attempts for its experimental realization have been made in various systems, including heavy fermion (HF) superconductors~\cite{matsuda-jpsj07}, organic superconductors~\cite{lortz-prl07} and cold atoms on optical lattices~\cite{liao-Nature09}. HF superconductors are promising due to their largely enhanced Pauli paramagnetic susceptibility. 

The tetragonal HF compound CeCoIn$_5$ undergoes a SC transition at 2.3~K at ambient pressure and zero magnetic field~\cite{petrovic:jpcm-01}. Power-law behavior in the low-temperature specific heat and thermal conductivity suggests an unconventional SC state~\cite{movshovich:prl-01}, while the extraordinary large electron mean-free path of $\sim$4000\AA~\cite{Seyfarth-prl08} sets SC in the clean limit. The Maki parameter, which characterizes the relative strength of Pauli to orbital limiting effects in magnetic field, ranges between 3.5 ($H\parallel [001]$) and 4.5 ($H\parallel [100]$)~\cite{bianchi:prl-02}, i.e. strongly exceeds the minimal value of 1.8, required for the FFLO state \cite{gruenberg:prl-66}. The discovery of an additional phase in the high-field and low-temperature (HFLT) corner of the SC phase diagram~\cite{bianchi:prl-03a,radovan:nature-03} was therefore proposed to be the realization of the long sought-after FFLO state and promoted numerous experimental and theoretical studies. Specific heat shows an anomaly across the transition between the low-field SC and the HFLT SC phases at magnetic fields above $H_{\rm HFLT}=10.3$~T along the tetragonal basal plane ($H_{\rm c2}=11.7$~T)\cite{bianchi:prl-03a,radovan:nature-03}. This HFLT SC phase was further confirmed by several measurement techniques~\cite{martin-prb05,capan-prb04,watanabe-prb04,correa-prl07,kakuyanagi-prl05,mitrovic-prl06,kumagai-prl06}. However, the formation of a FFLO phase in this material is controversial, since the subsequent NMR and neutron scattering studies found an incommensurate small-moment antiferromagnetic (AF) order in the HFLT SC state, which is likely of spin-density-wave (SDW) type~\cite{Young-PRL07,Kenzelmann-Science08}. One of the most peculiar properties of this phase is that the AF order does not extend into the normal state and exists only in the SC state~\cite{Kenzelmann-Science08}, suggesting some additional stabilization of AF order by the SC state. Some theories proposed mechanisms for stabilizing AF order due to strong Pauli-limiting and a nodal SC gap structure, without a FFLO state~\cite{Ikeda-prb10-2,Kato-prl11}, while in another theory, a coexisting FFLO state is necessary for the formation of AF order~\cite{Yanase-jpsj09}. A recent In-NMR study for fields along [100] has suggested the formation of a pure FFLO phase leading to an anomalous line broadening already at fields between 9.2~T and $H_{\rm HFLT}=10.3$~T, while the coexistence of AF order and FFLO superconductivity is claimed at $H_{\rm HFLT}\leq H\leq H_{\rm c2}$~\cite{Koutroulakis-prl10}. A spatially uniform coexistence of AF order and FFLO nodal planes has also been suggested from the most recent NMR study~\cite{Kumagai-prl11}. The ordered moment associated with the AF order disappears when the field is rotated by more than 17$^\circ$ out of the basal plane~\cite{blackburn-prl10}. Thus, any remaining anomalies, in particular for $H\parallel[001]$, could not be related to AF order.
The Maki parameter for this field direction is still twice as large as the required value for the formation of FFLO state, but much less studies on the SC state at high fields for $H\parallel[001]$ have been reported. NMR experiments~\cite{kumagai-prl06} suggest a FFLO state with a rather temperature independent phase boundary around 4.7~T, i.e. very close to $H_{\rm c2}$=5.0~T, in contrast to the strongly temperature dependent $H_{\rm HFLT}$ transition along the basal plane direction~\cite{bianchi:prl-03a}.

In the FFLO state, the SC gap function, $\Delta ({\bf r})$, is spatially modulated with a wave length, 2$\pi/q$, and paramagnetic quasiparticles appear periodically at nodal positions ($\Delta$=0). Because of the additional quasiparticles in the FFLO state above the critical field, $H_{\rm FFLO}$, the isothermal entropy as a function of field shows an additional convex $\sqrt{H-H_{\rm FFLO}}$-like contribution, leading to a steep increase of $S(H,T=const)$~\cite{Suzuki-private,PhysRevA.83.063621}. Any magnetic ordering, by contrast, will have a negative isothermal entropy contribution related to the reduction of degrees of freedom. Therefore, the observation of a steep {\it increase} would be a "smoking gun" proof of FFLO SC.
Experimentally, we can very precisely determine the isothermal field dependence of the entropy by measuring the magnetic Gr\"uneisen ratio $\Gamma_H=1/T(dT/dH)_S$, i.e., the magnetocaloric effect under perfect adiabatic conditions, and the specific heat $C$ using the thermodynamic relation 

\begin{eqnarray}
\left.\frac{dS}{dH}\right|_T=-\frac{C}{T}\left.\frac{dT}{dH}\right|_S . 
\end{eqnarray}

High quality single crystals were grown by the self-flux method. The specific heat and magnetic Gr\"uneisen ratio were measured with very high resolution in a dilution refrigerator with a SC magnet equipped with an additional modulation coil by utilizing heat-pulse and alternating field techniques, respectively~\cite{tokiwa-rsi11}. Using (1), we could resolve entropy changes as small as $2\times 10^{-5}$J/mol$\cdot$K within the SC state of CeCoIn$_5$, corresponding to 3.5~ppm of $R\log 2$.

We first concentrate on measurements for $H\parallel [100]$, shown in Figure 1. The overall convex shape of the field-dependence of the heat capacity at 0.2~K results from the strong Pauli limiting effect of the SC~\cite{Ichioka-prb07}. Pronounced peaks in the heat-capacity and magnetic Gr\"uneisen ratio indicate the first-order transition to the normal state at $H_{\rm c2}=11.5$~T. The additional second-order transition between the regular and HFLT-SC states at $H_{\rm HF}$ leads to broadened discontinuities in the two quantities. The temperature dependence of $H_{\rm HF}$  (cf. Inset Fig. 1b) is in perfect agreement with previous results~\cite{bianchi:prl-03a}. Using (1), we determine the field-dependence of the entropy at 0.2~K (cf. Fig. 1b). At  $H_{\rm HF}$=10.4~T, the field derivative $(dS/dH)_T$ displays a broadened downwards discontinuity indicating a {\it negative} contribution to the entropy. The overall increase of entropy with field is naturally explained by the increasing number of paramagnetic quasiparticles. We do not resolve a phase transition at 9.2~T, i.e. the field at which a drastic broadening of the In-NMR spectra has suggested "exotic superconductivity"~\cite{Koutroulakis-prl10}. This excludes the formation of a FFLO state in this part of the phase diagram.

\begin{figure}
\includegraphics[width=\linewidth,keepaspectratio]{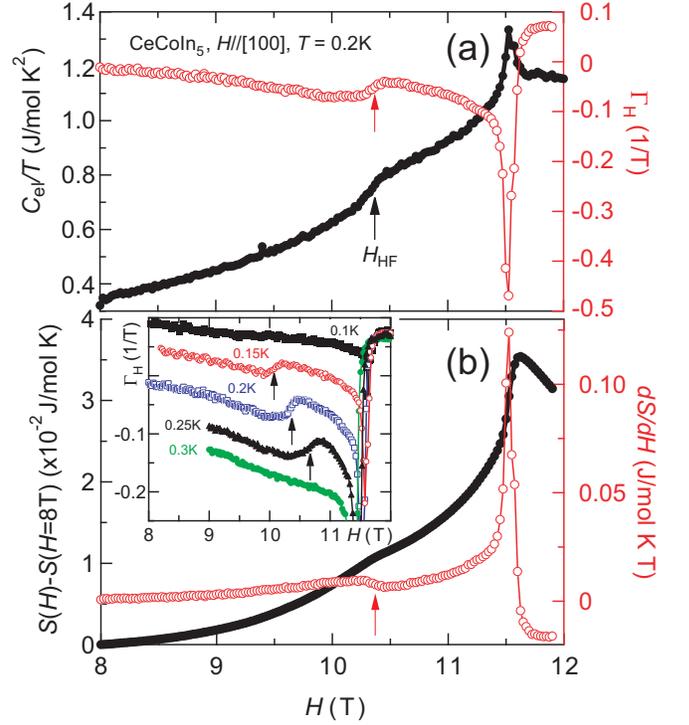}
\caption{(Color online) (a) Magnetic field dependence of electronic specific heat divided by temperature (black solid circles, left axis) and magnetic Gr\"{u}neisen ratio (red open circles, right axis) of CeCoIn$_5$ at 0.2~K for $H\parallel$[100]. (b) Calculated field-derivative of the entropy (red open circles, right axis) and integrated entropy increment (black solid circles, left axis).  Inset displays field-dependence of magnetic Gr\"{u}neisen ratio at various different temperatures. Arrows indicate the transition to the high-field SC phase, $H_{\rm HF}$.}
\end{figure}

\begin{figure}
\includegraphics[width=\linewidth,keepaspectratio]{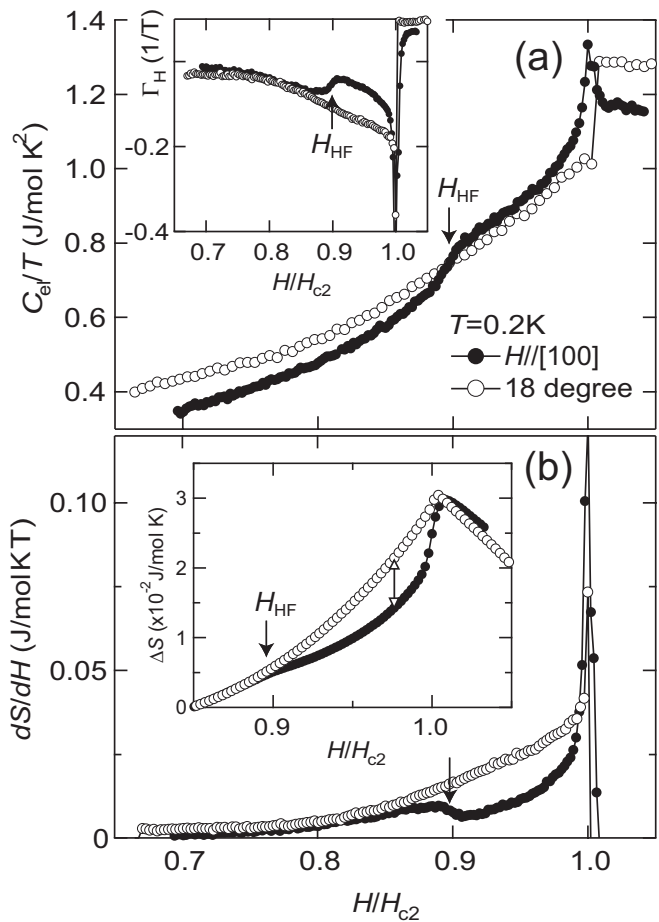}
\caption{(a) Electronic specific heat divided by temperature of CeCoIn$_5$ at 0.2K for $H\parallel[100]$ (solid circles) and 18$^{\circ}$ tilted from [100] towards [001] (open circles) versus magnetic field normalized by upper critical field $H_{\rm c2}=11.7$~T ($H\parallel[100]$) and 9.0~T ($H$ tilted by 18$^{\circ}$  towards [001]), respectively. Inset shows the respective magnetic Gr\"{u}neisen ratio data. (b) Calculated field-derivative of the entropy. Inset shows respective entropy increments for $H\parallel[100]$ (solid circles) and 18$^{\circ}$ (open circles). Solid arrows indicate the high-field transition for $H\parallel[100]$. Two-sided open arrow in inset indicates the entropy difference of 8\,mJ/mol$\cdot$K.}
\end{figure}

Since previous neutron diffraction measurements have proven that the AF state disappears at out-of-plane angles larger than 17$^\circ$~\cite{blackburn-prl10}, in accordance with recent theoretical work~\cite{Suzuki-prb11}, it is of particular interest to compare the measurements for $H\parallel[100]$ with respective measurements in tilted field, cf. Figure 2a. Indeed the thermodynamic signatures of the HFLT transition have disappeared at a tilting angle of 18$^{\circ}$ for which monotonic behavior up to $H_{\rm c2}$ is found in the various properties~\cite{correa-prl07}. Using the field dependence of $(dS/dH)_{T}$, we extract with high precision the isothermal field dependence of the entropy (inset of Fig.~2b). At $H<H_{HF}$ and $H>H_{c2}$ similar behavior is found for both field orientations, while in the intermediate field regime a depression is visible for $H\parallel [100]$. This negative entropy increment of 8~mJ/mol$\cdot$K at 0.2~K (cf. two-sided arrow in the inset) must be related to the AF ordering for $H\parallel[100]$, which disappears at 18$^{\circ}$. Using the value of the specific heat coefficient at the same temperature, $C_{\rm el}/T= 0.67$\,J/mol$\cdot$K$^2$, we estimate a 6\% reduction of the DOS at the Fermi energy. Chromium, for example as comparison, loses a similar fraction, $\sim$4\% of DOS due to the gapping of the Fermi surface at the SDW transition~\cite{overhauser-10}. The fraction of truncated Fermi surface area, $t$, can be estimated from the SDW modulation vector, $Q$=(0.56,0.56,0.5)~\cite{Kenzelmann-Science08}, and the SDW gap, $\Delta_{\rm SDW}$, as $t$=(2$p\Delta_{\rm SDW} m^\star$)/($\left|Q\right|^2\hbar^2$) where $p$ is the number of truncated faces and $m^\star$ the effective mass~\cite{overhauser-10}. The gap is approximated by the BCS formula, $\Delta_{\rm SDW}$=3.5$k_{\rm B}T_{\rm SDW}$, using the phase transition temperature of 0.27~K~\cite{bianchi:prl-03a}. For a tetragonal system $p=4$ and $m^\star$ can be estimated from the Sommerfeld coefficient, $\gamma$=0.67J/mol$\cdot$K$^2$, as $m^\star=(3\hbar^2\gamma)/(k_{\rm B}^2V_mk_{\rm F})$=230$m_0$ for a spherical Fermi surface and $m^\star=(3\hbar^2\gamma)/(k_{\rm B}^2a^2N_{\rm A})$=1000$m_0$ for a cylindrical Fermi surface. Here $V_m$ is molar volume, $k_{\rm F}=Q/2$, $a=4.62\AA$ and $N_{\rm A}$ denotes Avogadro's constant. Then, $t$ equals 0.015 and 0.065 for spherical and cylindrical Fermi surfaces, respectively. In the quasi-2D case relevant for CeCoIn$_5$ and intermediate value is expected, in good agreement with the experimentally observed $t=0.06$.

We also estimate the increase of the entropy for a possible FFLO state, which arises due to a spatial modulation of the SC gap along the direction of the applied field. The most energetically stable wavelength of the gap modulation in the FFLO state was theoretically determined as a function of magnetic field~\cite{Suzuki-jpsj11}. A long wave length right above $H_{\rm FFLO}$ decreases sharply as the field is increased and reaches the smallest value of 17\,$R_0$ at a magnetic field close to $H_{\rm c2}$, where $R_0$=($\hbar v_{\rm F}$)/($2\pi k_{\rm B}T_c$) and $v_{\rm F}$ denotes the Fermi velocity~\cite{Ichioka-prb07}. The reported in-plane $v_{\rm F}=7000$~m/s~\cite{Knebel-jpsj08,MicleaCF:PredFs}, yields $R_0$=38\AA \hspace{1ex} and the shortest modulation wavelength, 17\,$R_0$=650\AA. Using the SC gap $\Delta_0$=4.92K~\cite{movshovich:prl-01}, we can calculate the spatial dependence of the quasiparticle occupation $f$($z$)=[1+$\exp(E_{\rm k}(z)/k_{\rm B}T$)]$^{-1}$ along the z-direction (parallel to the field). Here $E_{\rm k}(z)=[(\epsilon_{\rm k}-E_F)^2+(\Delta_0\sin(2\pi z/17R_0))^2]^{1/2}$ and $\epsilon_{\rm k}$ denotes the kinetic quasiparticle energy. Note, that at the nodal positions $f$ equals the normal Fermi Dirac function, giving rise to an enhancement of the entropy compared to the SC state. The  position dependent entropy is derived from $S(z)=-2k_{\rm B}\Sigma_k[(1-f(z))\ln(1-f(z))+f(z)\ln(f(z))]$. Its z-average at 0.2~K amounts to 15\% of the entropy difference between $H_{HF}$ and $H_{c2}$. Using $S$($H_{\rm c2})-S(H_{\rm HF}$)=25mJ/mol$\cdot$K, this 15\% increment corresponds to 4~mJ/mol$\cdot$K. Consequently, within a SDW-FFLO coexistence scenario, this increase needs to be overcompensated by an entropy reduction due to the SDW ordering of 12~mJ/mol$\cdot$K at 0.2~K, in order to match the experimental data. This would correspond to $t\approx 0.09$, which is well beyond the estimated upper limit for the fraction of DOS which could be gapped due to the SDW formation. We therefore conclude that the data indicate a SDW ordering without FFLO state.

 
We now turn our attention to measurements in magnetic fields applied 90$^{\circ}$ off the basal plane, i.e. parallel [001]. Figure 3 shows the isothermal field dependences of $C_{el}/T$, $\Gamma_H$ and the corresponding $(dS/dH)_T$ and $\Delta S$ at 0.2~K. The overall concave shape of $C_{el}/T$ is similar as found for $H\parallel[100]$. A sharp peak in the field-derivative of the entropy at $H_{\rm c2}=4.9$~T is characteristic of the first order SC transition, caused by the strong Pauli limiting~\cite{bianchi:prl-02}. Whereas the isothermal field derivative of the entropy $(dS/dH)_T$ has shown a step-like decrease at $H\parallel [100]$ compatible with a second-order phase transition, for $H\parallel [001]$ it displays only a change in slope. Recent isothermal magnetization measurements have found related behaviors for the magnetic susceptibility $dM/dH$ along the two field directions\cite{Gratens-prb12}. Changes in slope of second-order derivatives of the free energy, such as $(dS/dH)_T$ correspond to discontinuities in third derivatives of free energy only, i.e. too weak for a second-order phase transition. With increasing temperatures, the observed kink broadens significantly and is shifted towards lower field values, as shown by the red squares in the inset of Fig.~3(a). $(dS/dH)_T$ for the field applied 70$^{\circ}$ tilted away from [100] towards [001] is also plotted in the inset of Fig.~3(b), to show the evolution of the kink with the field angle. It is still visible, however broadened. Presumably, this feature broadens further with tilting angle and smoothly changes to a featureless curve, similar to that for the field 18$^{\circ}$ off the basal plane.

The origin of this "kink-signature" could not be related to the previously observed AF ordering, which has shown to disappear at angles larger than  17$^{\circ}$~\cite{blackburn-prl10}. The formation of a FFLO state, suggested by earlier NMR experiments~\cite{kumagai-prl06}, could be excluded as well, since, as explained above, a sharp increase of $(dS/dH)_T$ upon increasing the field across $H_{\rm FFLO}$ due to a sudden increase of the quasiparticle entropy would have been expected. Another possibility would be that the anomaly is related to a change of the vortex lattice. Small angle neutron scattering has previously revealed a first-order rhombic to hexagonal transition of the vortex lattice near 4.4~T~\cite{bianchi-science08}. These experiments have detected also a second-order square to rhombic transition around 3.3~T, which our measurements do not detect. Observation of vortex lattice phase transitions by bulk measurements is very rare, except for solid-liquid transitions in cuprates, exhibiting entropic anomalies~\cite{Schilling-nature96,ZELDOV-nature95}. 

Previous Hall effect and thermal expansion measurements in the normal state of CeCoIn$_5$ for $H\parallel c$ have found crossover lines $T_{cr}(H)$ and $T_{FL}(H)$ separating non-Fermi liquid from Fermi liquid behavior, which display a linear temperature vs field relation and extrapolate to a critical field of roughly 4~T in the zero-temperature limit~\cite{Singh-prl07,Zaum-prl11}. This suggests a quantum critical point (QCP), hidden by the SC phase and possibly related to the suppression of AF order emerging under negative pressure or Cd-doping~\cite{Zaum-prl11,PhamLD:Revtth}. Generically, the magnetic Gr\"uneisen parameter and thereby $(dS/dH)_T$ are expected to display a characteristic sign change within the quantum critical regime, due to the accumulation of entropy~\cite{GarstM:SigctG}. For CeCoIn$_5$, the field dependence of the entropy in the SC state is dominated by the vortex-lattice quasiparticle contribution. Assuming a smooth evolution of the latter, as sketched by the dashed line in the inset of Fig.~3b, the observed kink-like anomaly may result from an additional quantum critical contribution which changes sign. Note, that quantum criticality at finite temperatures could result from a nearby QCP in multi-parameter space, e.g. being located at a finite Cd-doping~\cite{PhamLD:Revtth}. Finally, we note that the observed anomaly further broadens upon tilting the field from [001] towards [100] and disappears at 18$^{\circ}$ from the in-plane field orientation.


\begin{figure}
\includegraphics[width=\linewidth,keepaspectratio]{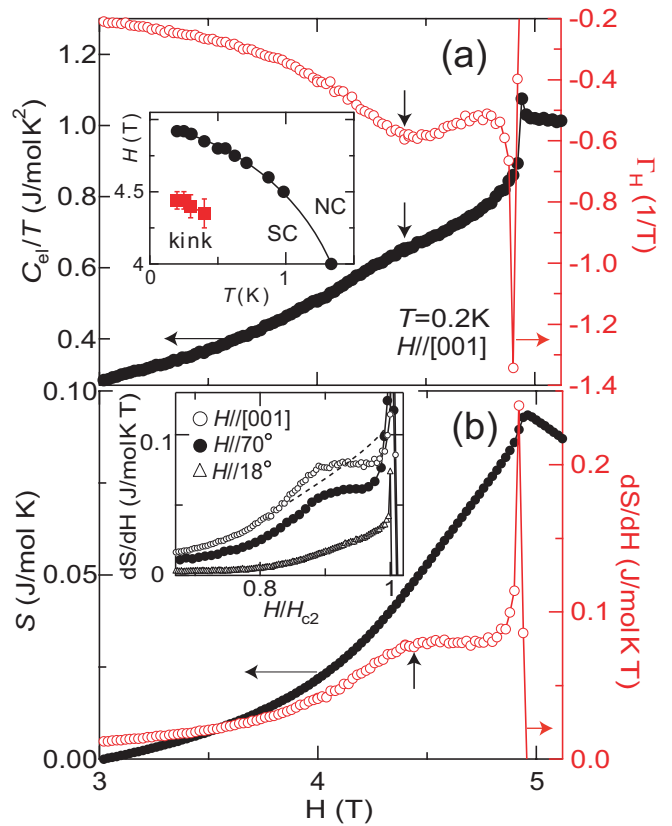}
\caption{(Color online) (a) Electronic specific heat divided by temperature (black solid circles, left axis) and magnetic Gr\"{u}neisen ratio (red open circles, right axis) of CeCoIn$_5$ as a function of field for $H\parallel[001]$ at 0.2\,K. Inset shows the superconducting phase diagram for $H\parallel[001]$. Black circles and red squares mark positions of upper critical field and "kink anomaly" (cf. black arrows in main panels), respectively. (b) Calculated field-derivative of the entropy, $(dS/dH)_T$, (red open circles, right axis) and integrated entropy increment (black solid circles, left axis). Inset shows $(dS/dH)_T$ for the field applied along [001] and different angles tilted away from [100] towards [001]. Dashed line see text.}
\end{figure}

In conclusion, our high-precision measurements of the isothermal field dependence of the entropy in the SC state of CeCoIn$_5$ do not find an additional component, which would indicate nodal quasiparticles in a FFLO SC state. By contrast, for isothermal measurements at $H\parallel [100]$ a clear {\it reduction} of the entropy by 8\,mJ/mol$\cdot$K at 0.2~K is found at a second-order HFLT transition at 10.4~T. This transition coincides with the previously detected incommensurate AF order~\cite{Young-PRL07,Kenzelmann-Science08}. The observed reduction of DOS is in perfect agreement with the expectation for a SDW formation without additional FFLO state. Upon tilting the field direction by 18$^{\circ}$ towards the $[001]$ direction, the HFLT transition has completely disappeared and a FFLO state could be excluded within the experimental resolution, which is three orders of magnitude better than the estimated increase of entropy due a FFLO state. A FFLO state could also be excluded for $H\parallel [001]$, where a broadened kink anomaly is found, which likely is related to a nearby QCP. Finally, we note that a similar study of the isothermal field dependence of the entropy could provide a conclusive test for the existence of FFLO SC states in other candidate materials such as organic superconductors~\cite{lortz-prl07,Cho-prb09,Tanatar-prb02}.

We are grateful to K. Machida, R. Movshovich, I. Vekhter and K. M. Suzuki for stimulating discussions. This work was supported by the German Science Foundation through FOR 960 (Quantum Phase Transitions). Work at Los Alamos National Laboratory was performed under the auspices of the U.S. Department of Energy, Office of Basic Energy Sciences, Division of Materials Sciences and Engineering.

\end{document}